\documentclass[12pt,a4paper]{article}
\usepackage{amsmath,amssymb,bbm} 
\usepackage{accents}
\usepackage{color}
\usepackage[bookmarks=true,hyperfigures=true]{hyperref}
\usepackage{graphicx}
\usepackage[nosort]{cite}
\usepackage[bulletsep]{collref}
\usepackage{tensor}

\usepackage{color}

\usepackage[a4paper,text={173mm,216mm},centering]{geometry}
\allowdisplaybreaks[3]

\usepackage[font=small,labelfont=bf,width=0.85\textwidth]{caption}

\let\oldbfseries=\bfseries
\let\oldmdseries=\mdseries
\let\oldnormalfont=\normalfont
\renewcommand{\bfseries}{\oldbfseries\boldmath}
\renewcommand{\mdseries}{\oldmdseries\unboldmath}
\renewcommand{\normalfont}{\oldnormalfont\unboldmath}

\makeatletter
\newlength{\apb@width}
\newcommand{\autoparbox}[2][c]{\settowidth{\apb@width}{#2}\parbox[#1]{\apb@width}{#2}}

\makeatother

\newcommand{\nn}{\nonumber}

\newcommand{\be}{\begin{equation}}
\newcommand{\ee}{\end{equation}}
\newcommand{\ba}{\begin{eqnarray}}
\newcommand{\ea}{\end{eqnarray}}
\ifx\genfrac\sdflkaj\else\fi
\newcommand{\sfrac}[2]{{\textstyle\frac{#1}{#2}}}
\newcommand{\half}{\sfrac{1}{2}}

\newcommand{\cA}{\mathcal{A}}

\newcommand{\cO}{\mathcal{O}}

\def\l<{\langle}\def\r>{\rangle}

\makeatletter
\newcommand{\namedref}[2]{\hyperref[#2]{#1~\ref*{#2}}}
\newcommand{\secref}{\@ifstar{\namedref{Section}}{\namedref{sec.}}}
\newcommand{\subsecref}{\@ifstar{\namedref{Subsection}}{\namedref{subsec.}}}
\newcommand{\appref}{\@ifstar{\namedref{Appendix}}{\namedref{app.}}}
\newcommand{\tabref}{\@ifstar{\namedref{Table}}{\namedref{tab.}}}
\newcommand{\figref}{\@ifstar{\namedref{Figure}}{\namedref{fig.}}}
\makeatother

\newif\ifmrnote 
\mrnotetrue
 
\usepackage[active]{srcltx}

\newif\ifjbnote 
\jbnotetrue
 
\newcommand{\eqn}[1]{(\ref{#1})}

\newcommand{\YM}{\text{YM}}
\newcommand{\GR}{\text{G}}

\newcommand{\derpa}[1]{\frac{\partial}{\partial p_a^{#1}}}
\newcommand{\sld}{S^{(1)}}
\newcommand{\ssld}{S^{(2)}}
\newcommand{\abr}[1]{\langle #1 \rangle}                                          % angle bracket
\newcommand{\sbr}[1]{[ #1 ]}                                                      % squarebracket
\newcommand{\ltil}[1]{\tilde{\lambda}^{\dot{#1}}}                                 % lambda tilde
\newcommand{\derl}[2]{\frac{\partial}{\partial \lambda_{#1}^{#2}}}                % der wrt lambda
\newcommand{\derlt}[2]{\frac{\partial}{\partial \tilde{\lambda}_{#1}^{\dot{#2}}}} % der wrt lambda
\newcommand{\cff}[1]{c^{(a)}_{#1}}                                                % coefficients for Omega
\newcommand{\cffb}[1]{\bar{c}^{(a)}_{#1}}                                         % coefficients for Omegabar
\newcommand{\se}{\epsilon}							  % soft epsilon

\begin{document}
\thispagestyle{empty}
% \ifarxiv\vspace*{-20mm}\fi

\begingroup\raggedleft\footnotesize\ttfamily
HU-EP-14/20\\
\vspace{15mm}
\endgroup

\begin{center}
{\LARGE\bfseries   Constraining subleading soft 
gluon \\ and graviton theorems
\par}%
\vspace{15mm}

\begingroup\scshape\large 
Johannes Broedel${}^{1}$, Marius de Leeuw${}^{1}$, \\[0.3cm] Jan Plefka${}^{1,2}$ and
Matteo Rosso${}^{1}$
\endgroup
\vspace{5mm}

\textit{${}^{1}$ Institut f\"ur Theoretische Physik,\\ Eidgen\"ossische
  Technische Hochschule
  Z\"urich,\\ Wolfgang-Pauli-Strasse 27, 8093 Z\"urich, Switzerland}\\[0.1cm]
  \texttt{\small \{jbroedel,deleeuwm,mrosso\}@itp.phys.ethz.ch\phantom{\ldots}} \\ \vspace{5mm}

\textit{${}^{2}$ Institut f\"ur Physik and IRIS Adlershof,\\ Humboldt-Universit\"at zu Berlin, \phantom{$^\S$}\\
  Newtonstra{\ss}e 15, D-12489 Berlin, Germany} \\[0.1cm]
\texttt{\small plefka@physik.hu-berlin.de\phantom{\ldots}} \vspace{8mm}

% \vspace{\fill}

\textbf{Abstract}\vspace{5mm}\par
\begin{minipage}{14.7cm}
  We show that the form of the recently proposed subleading soft graviton and
  gluon theorems in any dimension are severely constrained by elementary
  arguments based on Poincar\'e and gauge invariance as well as a
  self-consistency condition arising from the distributional nature of
  scattering amplitudes. Combined with the assumption of a local form as it
  would arise from a Ward identity the orbital part of the subleading operators
  is completely fixed by the leading universal Weinberg soft pole behavior. The
  polarization part of the differential subleading soft operators in turn is
  determined up to a single numerical factor for each hard leg at every order
  in the soft momentum expansion. In four dimensions, factorization of the
  Lorentz group allows to fix the subleading operators completely. 
\end{minipage}\par
\end{center}
\newpage

%%%%%%%%%%%%%%%%%%%%%%%%%%%%%
% \pagenumbering{arabic} \setcounter{page}{1}
% \renewcommand{\thefootnote}{\arabic{footnote}} \setcounter{footnote}{0}

% \setcounter{tocdepth}{2} \hrule height 0.75pt
% \tableofcontents
% \vspace{0.8cm} \hrule height 0.75pt \vspace{1cm}

\setcounter{tocdepth}{2}

%%%%%%%%%%%%%%%%%%%%%%%%%%%%%%%%%%%%%%%%%%%%%%%%%%%%%%%%%%%%%%%%%%%%%%%%%%%%%%%%
%%%%%%%%%%%%%%%%%%%%%%%%%%%%%%%%%%%%%%%%%%%%%%%%%%%%%%%%%%%%%%%%%%%%%%%%%%%%%%%%

\section{Introduction}

Gluon and graviton scattering amplitudes display a universal factorization
behavior when a gluon (respectively photon) \cite{Low:1958sn} or a graviton
\cite{Weinberg:1964ew} becomes soft, as was shown more than 50 years ago. This
leading soft pole behavior is known as Weinberg's soft theorem
\cite{Weinberg:1964ew,Weinberg:1965nx}. Recently an interesting proposal was
put forward by Cachazo and Strominger \cite{Cachazo:2014fwa} in which they
conjectured the extension of this theorem for gravitons to subleading and
sub-subleading orders in the soft momentum expansion. The proposal was shown to
hold at tree-level using the BCFW recursion relations \cite{Britto:2005fq}.
Tree-level gluon amplitudes exhibit a very similar subleading universal
behavior as pointed out in ref.~\cite{Casali:2014xpa} using a proof identical
to that of gravitons. In fact such a subleading gluon relation was argued to exist
already in refs.~\cite{Low:1958sn,Burnett:1967km}. Similarly, the
subleading soft graviton behavior was reported already in 1968
\cite{Gross:1968in}, see also the more recent discussion \cite{White:2011yy}.  

Collectively these (new) subleading
soft theorems state the existence of certain universal differential operators
in momenta and polarizations acting on a hard $n$-point amplitude, which
capture the subleading or even sub-subleading terms in the soft limit of the
associated $(n+1)$-point amplitude with one leg taken soft. For the case of
gravity the subleading soft theorems have been conjectured to be Ward
identities of a new symmetry of the quantum gravity $S$-matrix
\cite{Strominger:2013jfa,He:2014laa,Cachazo:2014fwa}, namely the extension of
the Bondi, van der Burg, Metzner and Sachs (BMS) symmetry
\cite{Bondi:1962px,Sachs:1962wk} to a Virasoro symmetry \cite{Barnich:2010eb}
acting on a sphere at past and future infinity.  This connection was first
established in ref.~\cite{He:2014laa} for the leading soft Weinberg pole term
\cite{Weinberg:1965nx}. Recently a
connection of the first subleading graviton theorem to the super-rotation
symmetry of extended BMS symmetry \cite{Barnich:2010eb} was reported
\cite{Kapec:2014opa}. Interesting steps towards a better understanding of such
a relation through dual holographic \cite{Adamo:2014yya} or ambitwistor
\cite{Geyer:2014lca} string models also appeared recently.

Inspired by these results a series of papers appeared
\cite{Casali:2014xpa,Schwab:2014xua,Bern:2014oka,He:2014bga,
Larkoski:2014hta,Cachazo:2014dia,Afkhami-Jeddi:2014fia,Schwab:2014fia,Bianchi:2014gla}.
Very interestingly the validity of the gluon and graviton subleading theorems
was shown to hold at \emph{any} dimension for tree-level amplitudes
\cite{Schwab:2014xua,Afkhami-Jeddi:2014fia}.  This is puzzling in the context
of the conjectured relation between gravity and extended BMS symmetry which is
clearly special to four dimensions. Similarly, it has been claimed in
\cite{Larkoski:2014hta} that the subleading soft theorem for gauge theory is
related to the conformal symmetry of tree-level gluon amplitudes, which again
contradicts the existence of the subleading theorem in general dimensions.

An important question is whether the subleading soft theorems receive radiative
corrections. Loop-level modifications of the leading soft-gluon theorem are
known to arise due to infrared singularities \cite{Bern:1998sc}. Whereas the
leading Weinberg soft graviton is protected, the subleading operators were
shown to be corrected in refs.~\cite{Bern:2014oka,He:2014bga}.
%In the recent works \cite{Bern:2014oka,He:2014bga} it was
%shown that while the leading Weinberg soft graviton theorem is protected all
%the subleading theorems do receive radiative corrections. 
This argument, however, was challenged in the recent work
\cite{Cachazo:2014dia}, where the authors argue for an order-of-limits problem:
Taking the soft limit prior to sending the dimensional regulator to zero would
not cause any corrections to the soft theorems.
%If one first takes the soft limit and then sends the
%dimensional regulator to zero there could be \emph{no} corrections to the soft
%theorems.

In this note we hope to shed some light on the above questions from a different
point of view.  We will show that rather elementary arguments can take one
quite far. Beyond the obvious Poincar\'e and gauge invariance we will assume a
certain \emph{local} form of the soft operators (as it would follow from a Ward
identity). In conjunction with a self consistency condition of the theorems
arising from the distributional nature of scattering amplitudes, the form of
the subleading operators is strongly constrained.
%Namely, assuming a certain \emph{local} form of the soft operators (as it would
%follow from a Ward identity) together with Poincar\'e and gauge invariance in
%conjunction with a self consistency condition of the theorems arising from the
%distributional nature of scattering amplitudes, strongly constrains the form of
%the subleading operators.  
Our argument applies to all dimensions and determines the orbital part of the
subleading operators uniquely from the form of the known leading pole
functions. 

While our argument does not prove the \emph{existence} of a universal
subleading soft gluon and graviton theorems, it states that if a such a
behavior exists, it is inevitably of the form proposed recently. Therefore the
only input needed from a potential new symmetry of the quantum gravity or gauge
theory S-matrix is the mere existence of a Ward identity pertaining to
subleading orders in the soft limit.  The form of the orbital part of the
theorems is then fixed -- at least at tree level.

This paper is organized as follows. In section 2 we provide our general
arguments and derive the central distributional constraint linking the
subleading operator in the soft theorems to the leading one. In section 3 we
apply the established constraints to the subleading soft operators for gluons
and gravitons and show that they are capable of fixing the orbital piece while
strongly constraining the polarization part. In section 4 we apply the same
reasoning to the sub-subleading soft graviton operator yielding identical
results. In section 5 we specialize to four dimensions and employ the
spinor-helicity formalism in order to find that the same line of arguments now
entirely determine the subleading soft operators. We end with a discussion in
section 6.

\section{General arguments}
\label{sec:framework}

Let us briefly summarize the subleading soft theorems and our central argument.
We will consider amplitudes in $D$-dimensional pure gauge and gravity theories
denoted by $\cA_{n}=\delta^{(D)}(P)\, A_{n}$, where $P=\sum_{a=1}^n p_a$ is the
total momentum. The soft momentum of leg $n+1$ is taken to be $\se\, q^{\mu}$,
which allows us to control the soft limit by sending $\se$ to zero. The
subleading soft theorems may be stated as\footnote{We only consider amplitudes
  where the external particles are of the same type.}
\begin{equation}
  \label{eq:softthm}
  \cA_{n+1}(p_{1},\ldots , p_{n}, \se\,q) = 
  \mathcal{S}^{[l]}(p_{1},\ldots , p_{n},\se\, q) \, \cA_{n}(p_{1},\ldots , p_{n}) +\cO(\se^{l}).
\end{equation}
where we call $\mathcal{S}^{[l]}$ a soft operator. The integer parameter $l$
controls the expansion in powers of the soft momentum to which the theorem
holds.

This theorem has been known to hold at leading order ($l=0)$ for more than 50
years. The corresponding soft factors in gauge theory \cite{Low:1958sn} and
gravity \cite{Weinberg:1964ew} read
\begin{equation}
  \mathcal{S}^{[0]}(\se\, q)= \begin{cases} \displaystyle \frac{1}{\se} S^{(0)}_\YM = \frac{1}{\se} \,
    \bigg(\frac{E\cdot p_{1}}{p_{1}\cdot q}-\frac{E\cdot p_{n}}{p_{n}\cdot q}\bigg) & \text{Yang--Mills theory (color ordered) }\\[12pt]  
    % \sum_{a=1,n; \text{signed}} \frac{E\cdot p_{a}}{p_{a} \cdot q} &
    % \text{Yang--Mills theory }\cr
    \displaystyle \frac{1}{\se} S^{(0)}_\GR = \frac{1}{\se} \,
    \sum_{a=1}^{n} \frac{E_{\mu\nu}\, p_{a}^{\mu}\, p_{a}^{\nu}}{p_{a} \cdot q}
    &\text{Gravity } \end{cases}
  \label{eq:S0explicit}
\end{equation}
where $E_{\mu}$ and $E_{\mu\nu}$ denote the gluon or graviton polarization of
the soft leg respectively and the arguments $\{p_{a}\}$ of
$\mathcal{S}^{[l]}(\se\, q)$ have been suppressed for brevity. Note that we are
working with color ordered gauge theory amplitudes\footnote{See
  e.g.~\cite{Henn:2014yza,Elvang:2013cua} for a textbook treatment.}. The soft
  limit is singular and the pole terms are universal. The graviton pole
  function $S^{(0)}_{\GR}$ does not receive radiative corrections
  \cite{Weinberg:1965nx,Bern:1998sv}.

In
refs.~\cite{Cachazo:2014fwa,Casali:2014xpa,Schwab:2014xua,Afkhami-Jeddi:2014fia}
the theorem in eqn.~\eqn{eq:softthm} has been demonstrated to extend to $l=1$ in
$D$-dimensional gauge theory and even $l=2$ in $D$-dimensional gravity at least
at tree-level
\begin{equation}
  \mathcal{S}^{[l]}(\se\, q)=
  \begin{cases}
    \displaystyle \frac{1}{\se} S^{(0)}_\YM
    +  S^{(1)}_\YM & \text{Yang-Mills theory } (l=1) \\[0.4cm]  \displaystyle
    \frac{1}{\se} S^{(0)}_\GR + S^{(1)}_\GR + \se \, S^{(2)}_\GR  &\text{Gravity }(l=2)\, .
  \end{cases}
\end{equation}
The operators $S_\YM^{(1)}$, $S_\GR^{(1)}$ and $S_\GR^{(2)}$ are differential
operators in the kinematical data of the hard legs and take a \emph{local}
form. Here with \emph{locality} we want to refer to the fact that they are sums
over terms depending on a single hard leg and the soft data only, i.e.
\begin{equation}
  S^{(l)}= \sum_{a} S^{(l)}_{a}(E,\se\,q;p_{a}, \partial_{p_{a}},E_{a},\partial_{E_{a}})\,.
  \label{eq:locality}
\end{equation}
This situation is just as one would expect it to arise from a Ward identity.

Naturally, the form of $\mathcal{S}^{[l]}$ is strongly restricted by Poincar\'e
and gauge invariance. While Poincar\'e invariance implies linearity in the
polarization tensors, gauge invariance demands vanishing of
$\mathcal{S}^{[l]}\cA_n$ order by order in $\se$ upon replacing the
polarizations by a gauge transformation.

There is, however, a further less obvious but elementary constraint on
$\mathcal{S}^{[l]}$ emerging from the distributional nature of amplitudes. The
left hand and the right hand side of the soft theorem eqn.~\eqn{eq:softthm}
depend on Dirac delta functions which differ in their arguments by the soft
momentum $\se \, q$. While this is no issue at leading order ($l=0$), it
becomes relevant for the subleading corrections. Therefore, in order for the
subleading soft theorems to be consistent, we need to require that
\begin{equation}
  \mathcal{S}^{[l]}(\se\, q)\, \delta^{D}(P) =
  \delta^{D}(P+ \se\, q) \, \tilde{\mathcal{S}}^{[l]}(\se\,q)
  \, ,
  \label{eq:conscond}
\end{equation}
where the soft operator $\tilde{\mathcal{S}}^{[l]}(\se\,q)$ acting on the
reduced amplitude $A_{n}$ could differ a priori from the soft operator
$\mathcal{S}^{[l]}(\se\,q)$ acting on the full amplitude $\cA_{n}$.
Interestingly, the results reported in the literature so far indicate that the
$\tilde{\mathcal{S}}^{[l]}(\se\,q)$ and $\mathcal{S}^{[l]}(\se\,q)$ are
equivalent\footnote{See in particular \cite{Bern:2014oka,Cachazo:2014dia} for a
discussion of different prescriptions related to this issue.}. We shall
show that this has to be the case.

Distinguishing different orders of $\se$, the soft theorem
eqn.~\eqn{eq:softthm} implies \cite{Cachazo:2014fwa} the relations
\begin{subequations}
  \label{eq:st}
  \begin{align}
    \lim_{\se\to 0}\, \Bigl (\se \,\cA_{n+1}(\se) \Bigr ) & = S^{(0)}\,
    \cA_{n}
    \label{eq:st1}\\
    \lim_{\se\to 0} \, \Bigl (\,\cA_{n+1}(\se) -\frac{1}{\se}\, S^{(0)}
    \cA_{n}\Bigr ) & = S^{(1)}\, \cA_{n}
    \label{eq:st2} \\
    \lim_{\se\to 0} \, \Bigl (\frac{1}{\se} \,\cA_{n+1}(\se)
    -\frac{1}{\se^{2}}\, S^{(0)} \cA_{n}-\frac{1}{\se}\, S^{(1)}\, \cA_{n}
    \Bigr ) & = S^{(2)}\, \cA_{n} \, .
    \label{eq:st3}
  \end{align}
\end{subequations}
In order to derive the implications of eqns.~\eqn{eq:st} for the soft operators
$S^{(0)}$ and $S^{(1)}$, it is useful to Laurent expand both the reduced
amplitude $A_{n+1}$ as well as its associated delta function:
\begin{align}
  A_{n+1}(\se) & = \frac{1}{\se}A_{n+1}^{(-1)}+A_{n+1}^{(0)}+\se A_{n+1}^{(1)}+\cO(\se^2)\\
  \delta^{(D)}(P + \se\,q) & = \delta^{(D)}(P) + \se (q\cdot\partial)\,
  \delta^{(D)}(P) +\cO(\se),
  \label{eq:Laurent1}
\end{align}
where we introduced the shorthand notation $q\cdot\partial = q^\mu
\frac{\partial}{\partial P^\mu}$. Let us now substitute these expansions into
eqns.~\eqn{eq:st1} and \eqn{eq:st2}. After noting that
\begin{equation}
  \label{eq:cconstr0}
  [S^{(0)} , \delta^{(D)}(P) ] =0\,,
\end{equation}
due to the form of $S^{(0)}$ in eqn.~\eqn{eq:S0explicit} being a mere function,
one finds for the \emph{reduced} amplitude from \eqn{eq:st2}
\begin{equation}
  \lim_{\se\to 0}A_{n+1}\,(\se)= \frac{1}{\se}\,
  S^{(0)}_\YM\,
  A_{n} +  A^{(0)}_{n+1} +\cO(\se)\,,
\end{equation}
For now, we will leave the form of the subleading contribution $A^{(0)}_{n+1}$
undetermined.  Eqn.~\eqref{eq:st1} then simply implies
\begin{equation}
  \label{eq:m3g}
  A_{n+1}^{(-1)} = S^{(0)} A_n \ .
\end{equation}
Eqn.~\eqref{eq:st2} leads to
\begin{align}
  S^{(1)} \cA_n & = \lim_{\se\to 0} \Bigl[ \cA_{n+1} - \frac{1}{\se}
  S^{(0)} \cA_n \Bigr] \ ,
  \notag\\
  & = \lim_{\se\to 0} \Bigl[ \Bigl( \delta^{(D)}(P) + \se
  (q\cdot\partial)\,\delta^{(D)}(P) \Bigr) \Bigl( \frac{1}{\se}
  A_{n+1}^{(-1)} + A_{n+1}^{(0)} \Bigr) - \frac{1}{\se} S^{(0)} \cA_n
  \Bigr] \ ,
\end{align}
where we kept only terms not vanishing as $\se\to 0$. Now we can remove the
limit on the right-hand side, whereas in the left-hand side we can commute
$S^{(1)}$ past the delta function to obtain
\begin{equation}
  \label{eq:stlast}
  \bigl[ S^{(1)}, \delta^{(D)}(P)\bigr] A_n +
  \delta^{(D)} (P) S^{(1)} A_n =
  \delta^{(D)}(P) A_{n+1}^{(0)} +  S^{(0)} \bigl( (q\cdot\partial) \delta^{(D)}(P)
  \bigr) A_n \ .
\end{equation}
At this point, several comments are in order. Most importantly, $\delta$ and
$\delta'$ may be treated as independent distributions if one takes partial
integration identities into account. Therefore we will have to match their
respective coefficients in order for this equation to be satisfied. Next,
$S^{(1)}$ must be a differential operator in the momenta $p_{a}$;
eqn.~\eqref{eq:stlast} implies then that
\begin{align}
  \label{eq:cconstr1}
  \bigl[S^{(1)},\delta^{(D)}(P)\bigr] &= S^{(0)}\,
  \Bigl( q\cdot\partial\,\delta^{(D)}(P)\Bigr)  + \chi \, \delta^{(D)}(P)\ , \\
  \label{eq:match1}
  A_{n+1}^{(0)} &= (S^{(1)} - \chi )\, A_n\ ,
\end{align}
where $\chi$ is an undetermined function.  Repeating the analysis for
eqn.~\eqref{eq:st3} (extracting the singular behavior from the reduced
amplitude, expanding in $\se$ and matching coefficients of the delta
function and its derivatives) leads to
\begin{align}
  \label{eq:cconstr2}
  \bigl[ S^{(2)}, \delta^{(D)}(P)\bigr] &= \tfrac{1}{2} \,S^{(0)}\, \Bigl(
  \bigl( q\cdot\partial \bigr)^2 \delta^{(D)}(P)\biggr) +  \Bigl(
  q\cdot\partial
  \delta^{(D)}(P) \Bigr)\, S^{(1)} + \chi' \, \delta^{(D)}(P)\,, \\
  \label{eq:s2geq}
  A_{n+1}^{(1)} &= (S^{(2)} - \chi') \, A_n \, .
\end{align}
We see that, the above equations constrain the subleading soft terms by relating
their form to the leading soft function $S^{(0)}$.  We will refer to those
equations as \emph{distributional constraints}.  Note also that the difference
of the soft operators ${\mathcal{S}}^{[l]}(\se\,q)$ and
$\tilde{\mathcal{S}}^{[l]}(\se\,q)$ mentioned in \eqn{eq:conscond} is captured
by -- a priori -- arbitrary functions $\chi$ and $\chi'$.

It is clear that the distributional constraints can only constrain the part of
$S^{(l)}$ that contains the derivatives with respect to the hard momenta.  We
call this piece the orbital part of $S^{(l)}$ and write
\begin{equation}
  S^{(l)}=  S^{(l)}_{\text{orb}} + S^{(l)}_{\text{polar}} +S^{(l)}_{\text{function}}
\end{equation}
with the orbital part
\begin{equation}
  S^{(l)}_{\text{orb}}= \sum_{a}
  S^{(l)\, \mu_{1}\ldots\mu_{l}}_{a}(E,q;p_{a})\, \frac{\partial}{\partial p_{a}^{\mu_{1}}}\ldots\frac{\partial}{\partial p_{a}^{\mu_{l}}}
\end{equation}
and the polarization part $S^{(l)}_{\text{polar}}$ containing derivatives
w.r.t.~the polarizations $E_{a}$. Finally
$
S^{(l)}_{\text{function}}
$
is a pure function of the soft and hard momenta linear
in the soft polarization $E$. It is not constrained by the 
distributional constraint as it commutes
with the Dirac delta function. 

As we are going to show below, distributional constraints, Poincar\'e, gauge
invariance and the assumption of locality of $S^{(l)}$ completely determine the
orbital part $S^{(l)}_{\text{orb}}$ of the soft operators in gauge theory and
gravity in any dimensions.  We now give a simple argument how to constrain also
the remaining polarization part.

In order to treat gluon and graviton polarizations on an equal footing let us
agree upon rewriting the graviton polarization of leg $a$ as
\begin{equation}
  E_{a\, \mu\nu} \to E_{a\, \mu}\, E_{a\, \nu} \qquad
  \text{with} \quad E_{a}\cdot E_{a}=0=p_{a}\cdot E_{a} \, .
\end{equation}
In four dimensions this no restriction at all, in general dimensions it is a
formal agreement which we can always undo at any stage due to the fact that an
amplitude is linear in the polarizations of all its legs. This replacement
unifies gauge and gravity theory in the sense that the same operators act on
the polarization degrees of freedom in both theories.  Using this prescription,
the operator representing a gauge transformations on leg $a$ takes the form
\begin{equation}
  W_{a}:= p_{a}\cdot \frac{\partial}{\partial E_{a}}
\end{equation}
and the Lorentz generators are represented as
\begin{equation}
  \label{LorentzpaEa}
  J^{\mu\nu} =\sum_{a} p_{a}^{\mu}\,  \frac{\partial}{\partial p_{a\,\nu}} + 
  E_{a}^{\, \mu}\,  \frac{\partial}{\partial E_{a\, \nu}} - \mu \leftrightarrow \nu\, 
\end{equation}
in both theories in any dimension\footnote{Strictly speaking this operator does
not generate the correct infinitesimal Lorentz transformation rule for the
polarizations as these do not transform as vectors, see
e.g.~\cite{Weinberg:1964ew,Weinberg:1995mt}. Next to the vector transformation
law there is an additional piece proportional to a gauge transformation in the
form of $W_{a}$. As this additional piece vanishes acting on amplitudes, the
form of \eqn{LorentzpaEa} is effectively correct.}. In this language the
polarization part $S^{(l)}_{\text{polar}}$ depends on the differential
operators $E_{a\, \mu}\, \frac{\partial}{\partial E_{a\, \nu}}$ in order to
preserve linearity of the amplitude in the polarization $E_{a}$.

Let us now consider gauge invariance of a fixed hard leg $a$ for the soft theorem
eqn.~\eqn{eq:softthm}
\begin{equation} 0 = W_{a}\, \cA_{n+1}(p_{1},\ldots , p_{n},\se\,q)= W_{a} \Bigl
  ( \mathcal{S}^{[l]}(\se\, q) \, \cA_{n}(p_{1},\ldots , p_{n}) \Bigr ) =
  \Bigl [ W_{a} , \mathcal{S}^{[l]}(\se\, q)\Bigr ] \, \cA_{n} =0\,.
  \label{Waconstraint}
\end{equation}
The orbital part $S^{(l)}_{\text{orb}}$ does not commute with $W_{a}$ due to the
presence of operators $\frac{\partial}{\partial p_{a}}$. Therefore it needs to be
completed to a gauge invariant structure. Employing the commutators
\begin{equation}
  \Bigl [W_{a},  p_{a}^{\mu}\frac{\partial}{\partial p_{a}^{\nu}}\Bigr] = - p_{a}^{\mu}\frac{\partial}{\partial E_{a\,\nu}}\, , \qquad
  \Bigl [W_{a},  E_{a}^{\mu}\frac{\partial}{\partial E_{a}^{\nu}}\Bigr] = + p_{a}^{\mu}\frac{\partial}{\partial E_{a\,\nu}}
  \, .
\end{equation}
the unique linear differential operator in $p_{a}$ and $E_{a}$ commuting with
$W_{a}$ reads
\begin{equation}
\label{Lambdadef}
  \Lambda_{a}^{\mu\nu}:= p_{a}^{\mu}\,  \frac{\partial}{\partial p_{a\, \nu}} + 
  E_{a}^{\, \mu}\,  \frac{\partial}{\partial E_{a\, \nu}}\, ,
\end{equation}
which we shall use as building block in constraining
$\mathcal{S}^{[l]}$ below\footnote{Note that in fact we only need the weaker
  condition of $[W_{a}, \mathcal{S}^{[l]}(\se\, q)\Bigr ] \sim W_{a}$ in
  eqn.~\eqn{Waconstraint} as $W_{a}$ annihilates the amplitudes $\cA_{n}$.  This is
  achieved by the operator $E_{a}\cdot\frac{\partial}{\partial E_{a}}$ which
  obeys
  $ [W_{a}, E_{a}\cdot\frac{\partial}{\partial E_{a}}] = W_{a} \, .  $
  However, as any amplitude is an eigenstate of the operator
  $E_{a}\cdot\frac{\partial}{\partial E_{a}}$ with eigenvalue one, including
  this operator in an ansatz for $\mathcal{S}^{[l]}$ is tantamount to writing a
  function.  We may therefore discard it in our analysis as functions of the
  kinematical data cannot be constrained. }.  Let us now turn to the explicit
analysis.

%%%%%%%%%%%%%%%%%%%%%%%%%%%%%%%%%%%%%%%%%%%%%%%%%%%%%%%%%%%%%%%%%%%%%%%%%%%%%%%%
\section{Subleading soft operators}

In this section we will apply the general framework outlined in the previous
section to determine the subleading soft operators in both gauge theory and
gravity.  As derived in \secref{sec:framework} above, the subleading
contribution should be fixed upon requiring locality, the distributional
constraint and gauge invariance for the soft leg. The last two requirements
translate into
\begin{align}
  \label{eq:s1com}
  &\bigl[\sld, \delta^{(D)}(P)\bigr] = S^{(0)} (q\cdot\partial) \delta^{(D)}(P)
  + \,\chi\, \delta^{(D)}(P) \ ,
  \\
  \label{eq:s1gauge}
  % &\sld (E \sim q) \cdot \cA_n = 0 \ .
  & \bigl[\sld, q\cdot\frac{\partial}{\partial E} \bigr]\cdot\mathcal{A}_n = 0 \
  .
\end{align}

\paragraph{Gauge theory.}
In gauge theory the leading order soft factor is given by the universal Weinberg
soft gluon function \cite{Weinberg:1965nx}
\begin{equation}
  S^{(0)}_\YM
  % = \sum_{a=n,1,\text{signed}}\frac{E_{\mu}\, p_{a}^{\mu}}{p_{a}\cdot q}
  =
  \frac{p_1 \cdot E}{p_1\cdot q} -
  \frac{p_n \cdot E}{p_n\cdot q}
  \label{S0ymrep}
\end{equation}
with the polarization vector $E_{\mu}$ for the soft particle.  We begin with an
ansatz for $S^{(1)}_{\text{YM}}$ reflecting the reasoning in
\secref{sec:framework}
\begin{equation}
  S^{(1)}_{\text{YM}}= \sum_{a} E_{\mu}\, \Omega_{a}^{\mu\nu\rho}\, 
  \Bigl( p_{a}^{\nu}\,  \frac{\partial}{\partial p_{a\, \rho}} + 
  E_{a}^{\, \nu}\,  \frac{\partial}{\partial E_{a\, \rho}}\Bigr ) \, .
\end{equation}
Before imposing the constraints \eqn{eq:s1com} and \eqn{eq:s1gauge} we note
that dimensional analysis and soft scaling requires $\Omega^{\mu\nu\rho}$ to be
of mass dimension -1 and to be scale invariant w.r.t.~the soft momentum $q$
respectively. In conjunction with the assumption of locality and $E\cdot
q=E_{a}\cdot p_{a}=p_{a}^{2}=0$ we are left with the compact ansatz
\begin{equation}
\label{ansatzomega}
  \Omega_{a}^{\mu\nu\rho}= c_{1}^{(a)}\, \frac{p_{a}^{\mu}\, q^{\nu}\, q^{\rho}}{(q\cdot p_{a})^{2}}
  + c_{2}^{(a)}\, \frac{\eta^{\mu\nu}\, q^{\rho}}{q\cdot p_{a}} 
  + c_{3}^{(a)}\, \frac{\eta^{\mu\rho}\, q^{\nu}}{q\cdot p_{a}} 
\end{equation}
where the numbers $c_{i}^{(a)}$ are to be determined. In order to do so we first
impose gauge invariance via \eqn{eq:s1gauge} which leads to
\begin{equation}
  0=\sum_{a}(c_{1}^{(a)}+c_{2}^{(a)}+c_{3}^{(a)})\, \frac{q^{\nu}\, q^{\rho}}{q\cdot p_{a}}\,
  \Bigl( p_{a}^{\nu}\,  \frac{\partial}{\partial p_{a\, \rho}} + 
  E_{a}^{\, \nu}\,  \frac{\partial}{\partial E_{a\, \rho}}\Bigr ) \, \cA_{n} \, .
\end{equation}
There is no way for this term to conspire to yield a Lorentz charge. Hence we
conclude that $c_{1}^{(a)}+c_{2}^{(a)}+c_{3}^{(a)}=0$. Turning to the
distributional constraint \eqn{eq:s1com} one now easily establishes
\begin{align} [S^{(1)}_{\YM},\delta^{(D)}(P)] &= \Bigl (\sum_{a} c_{3}^{(a)}\Bigr)\,
  E\cdot \partial\delta^{(D)}(P)
  - \Bigl (\sum_{a} c_{3}^{(a)}\,\frac{E\cdot p_{a}}{q\cdot p_{a}}\Bigr )\, q\cdot\partial\delta^{(D)}(P)\nn\\
  &\stackrel{!}{=} \Bigl ( \frac{p_1 \cdot E}{p_1\cdot q} - \frac{p_n \cdot E}{p_n\cdot q}
  \Bigr ) \, q\cdot\partial\delta^{(D)}(P) + \chi\, \delta^{(D)}(P)\, ,
\end{align}
where we have inserted eqn.~\eqn{S0ymrep} for $S^{(0)}_{\text{YM}}$. Solving for the undetermined coefficients, we find 
\begin{equation}
  c_{3}^{(1)}=-1\, , \quad c_{3}^{(n)}=1\, , \quad c_{3}^{(a)}=0 \, \quad \text{for }a=2,\ldots,n-1\, ,
\end{equation}
along with the vanishing of $\chi$, which implies the identity
$\mathcal{S}^{[1]}= \tilde{\mathcal{S}}^{[1]}$ (cf. eqn.~\eqn{eq:conscond}).
As $c_{3}^{(a)}= -c_{1}^{(a)}-c_{2}^{(a)}$ the
differences of the remaining coefficients $c_{-}^{(a)}:=c_{1}^{(a)}-c_{2}^{(a)}$
remain unconstrained. In fact they only couple to the polarization degrees of
freedom, the orbital part of $S^{(1)}$ is completely determined. 
In summary we have established that
\begin{align}
  S^{(1)}_{\YM} &= \sum_{a=1,n,\text{signed}} \frac{E_{\mu}q_{\nu}}{p_{a}\cdot
    q}\, \Bigl ( p_{a}^{\mu}\, \frac{\partial}{\partial p_{a\, \nu}} +
  E_{a}^{\, \mu}\,  \frac{\partial}{\partial E_{a\, \nu}} - \mu\leftrightarrow \nu \Bigr ) \nn\\
  & \quad + \sum_{a} \tilde c^{(a)}\, \Bigl ( \frac{(E\cdot p_{a})(E_{a}\cdot
    q)}{p_{a}\cdot q} - E\cdot E_{a} \Bigr ) \frac{1}{p_{a}\cdot q}\, q\cdot
  \frac{\partial}{\partial E_{a}}\, ,
\end{align}
where the undetermined coefficients $\tilde c^{(a)}$ are related to the
previous ones via $\tilde c^{(1)}= c^{(1)}_{-}+\tfrac 12$, $\tilde c^{(n)}=
c^{(n)}_{-}-\tfrac 12$ and $\tilde c^{(a)}= c^{(a)}_{-}$ for $a=2,\ldots, n-1$.
Note that the second sum is manifestly gauge invariant with respect to the soft
and hard legs.  Hence, the orbital part of the subleading soft operator is
entirely determined by our constraints and coincides with the explicit
tree-level computations in the literature.  The polarization piece is
constrained up to a single numerical factor for every hard leg.

Finally, let us briefly comment on the possible functional contribution $
S^{(1)}_{\YM\, \,\text{function}}$ at the subleading level. Assuming locality and
dimensional arguments quickly rules out any contribution here.

\paragraph{Gravity.}
The analysis of the graviton soft operator is almost a carbon copy of the gauge theory
one. The leading universal soft function for gravitons reads
\cite{Weinberg:1965nx}
\begin{equation}
  S^{(0)}_\GR= \sum_{a=1}^{n}\frac{E_{\mu\nu}\, p_{a}^{\mu}\,p_{a}^{\nu}
  }{q\cdot p_{a}} \, .
  % \frac{\vev{n1}}{\vev{ns}\vev{s1}}\, .
\end{equation}
We again start with an ansatz for $S^{(1)}_\GR$ of the form
\begin{equation}
  S^{(1)}_{\text{G}}= \sum_{a} E_{\mu\nu}\, \Omega_{a}^{\mu\nu\rho\sigma}\, 
  \Bigl( p_{a}^{\rho}\,  \frac{\partial}{\partial p_{a\, \sigma}} + 
  E_{a}^{\, \rho}\,  \frac{\partial}{\partial E_{a\, \sigma}}\Bigr ) \, .
  \label{S1gansatz}
\end{equation}
Dimensional analysis requires $\Omega_{a}^{\mu\nu\rho\sigma}$ to be of mass
dimension zero and to be scale invariant w.r.t.~the soft momentum $q$. This
together with the assumption of locality and the relations
$E_{\mu\nu}q^{\nu}=E_{a}\cdot p_{a}=p_{a}^{2}=0$ leads us to the most general
ansatz
\begin{equation}
\label{ansatzomega2}
  \Omega_{a}^{\mu\nu\rho\sigma}= c_{1}^{(a)}\, \frac{p_{a}^{\mu}\, p_{a}^{\nu}\, q^{\rho}
    \, q^{\sigma}}
  {(q\cdot p_{a})^{2}} + c_{2}^{(a)}\, \frac{\eta_{\phantom\mu}^{\rho(\mu}p_{a}^{\nu)}q^{\sigma}}
  {q\cdot p_{a}}+  c_{3}^{(a)}\, \frac{\eta_{\phantom\mu}^{\sigma(\mu}p_{a}^{\nu)}q^{\rho}}
  {q\cdot p_{a}} + c_{4}^{(a)}\, \eta^{\rho(\mu}\, \eta^{\nu)\sigma}\, ,
\end{equation}
with four undetermined numerical coefficients for each hard leg $a$. Imposing
gauge invariance for the soft leg amounts to the replacement $E_{\mu\nu}\to
\Lambda_{(\mu}q_{\nu)}$ in \eqn{S1gansatz}. We then obtain the condition
\begin{align}
  0&=\tfrac 1 2 \sum_{a}\Bigl [ (2c_{1}^{(a)}+c_{2}^{(a)}+c_{3}^{(a)})\,
  q^{\rho}\, q^{\sigma}\, \frac{\Lambda\cdot p_{a}} {q\cdot p_{a}} \nn\\ &\qquad
  + (c_{3}^{(a)}+c_{4}^{(a)})\, q^{\rho}\, \Lambda^{\sigma} +
  (c_{2}^{(a)}+c_{4}^{(a)})\, q^{\sigma}\, \Lambda^{\rho} \Bigr ]\, \Bigl(
  p_{a}^{\rho}\, \frac{\partial}{\partial p_{a\, \sigma}} + E_{a}^{\, \rho}\,
  \frac{\partial}{\partial E_{a\, \sigma}}\Bigr ) \, \cA_{n}\, .
\end{align}
The first term in the above requires
$2c_{1}^{(a)}+c_{2}^{(a)}+c_{3}^{(a)}=0$. For the second and third term we have
to be somewhat more careful. Here we have the possibility of these two terms
conspiring to build up the total Lorentz generator $J^{\rho\sigma}$ of
\eqn{LorentzpaEa} which annihilates $\cA_{n}$. We thus require
\begin{equation}
  c_{3}^{(a)}+c_{4}^{(a)}= - c_{2}^{(a)}-c_{4}^{(a)} = c
\end{equation}
with a \emph{universal} constant $c$ identical for all hard legs. We now move on
to pose our distributional constraint \eqn{eq:s1com} linking
$S^{(1)}_{\text{G}}$ to $S^{(0)}_{\text{G}}$. One finds
\begin{align} [S^{(1)}_{\text{G}}, \delta^{(D)}(P)] &=
  \sum_{a}(c_{1}^{(a)}+c_{2}^{(a)}) \,
  \frac{(p_{a}^{\mu}p_{a}^{\nu}E_{\mu\nu})}{q\cdot p_{a}}
  q\cdot\partial\delta^{(D)}(P)+ c\, P^{\mu}\, E_{\mu}{}^{\nu}\,
  \frac{\partial}{\partial P^{\nu}}\delta^{(D)}(P)
  \nn\\
  &\stackrel{!}{=} S^{(0)}_{\text{G}}\, q\cdot\partial\delta^{(D)}(P) + \chi \,
  \delta^{(D)}(P)\, .
\end{align}
One nicely sees that the first term on the r.h.s.~of the first line forms the
leading Weinberg soft function for the uniform choice
\begin{equation}
  c_{1}^{(a)}+c_{2}^{(a)}=1\, .
\end{equation}
The following term vanishes in the distributional sense by the tracelessness of
$E_{\mu\nu}$.  And finally we again learn that the function $\chi=0$ implying
again the identity of $\mathcal{S}^{[1]}= \tilde{\mathcal{S}}^{[1]}$ in the
sense of eqn.~\eqn{eq:conscond}. The established three equations for the four
unknowns may now be solved upon expressing everything in terms of $c_{4}^{(a)}$
\begin{equation}
  c_{1}^{(a)}=c_{4}^{(a)}\, , \qquad c_{2}^{(a)}=1-c_{4}^{(a)}\, , \qquad
  c_{3}^{(a)}=-1-c_{4}^{(a)}\, .
\end{equation}
One also checks that $c_{3}^{(a)}+ c_{4}^{(a)}=1$ in line with the above
reasoning.  Inserting this into the ansatz \eqn{S1gansatz} yields the final result
\begin{align}
  S^{(1)}_{\text{G}} &= \sum_{a=1}^{n} \frac{(p_{a}\cdot E)
    E_{\rho}q_{\sigma}}{p_{a}\cdot q}\, \Bigl ( p_{a}^{\rho}\,
  \frac{\partial}{\partial p_{a\, \sigma}} + E_{a}^{\, \rho}\,
  \frac{\partial}{\partial E_{a\, \sigma}} - \rho\leftrightarrow \sigma
  \Bigr ) \nn\\
  & \quad + \sum_{a=1}^{n} \tilde c^{(a)}\, \Bigl ( \frac{(E\cdot
    p_{a})(E_{a}\cdot q)}{p_{a}\cdot q} - E\cdot E_{a} \Bigr ) \Bigr [
  \frac{p_{a}\cdot E}{p_{a}\cdot q}\, q\cdot \frac{\partial}{\partial E_{a}}
  -E\cdot\frac{\partial}{\partial E_{a}}\, \Bigr ] \, ,
  \label{eq:s1final}
\end{align}
where we have renamed $c_{4}^{(a)}=\tilde c^{(a)}$ and written the soft
polarization $E_{\mu\nu}\to E_{\mu}\, E_{\nu}$ for compactness of notation.

We thus see that again the orbital part is completely determined and coincides
with the results established in the literature for tree-level amplitudes
\footnote{In fact it is in accordance with the expression for $\sld_\GR$ given
  in ref.~\cite{He:2014bga} and differs by an overall normalization factor in the expression
  of ref.~\cite{Cachazo:2014fwa}.}. The polarization-dependent parts are
constrained to one numerical factor for every hard leg, just as it was the case
in gauge theory.

Finally, let us also comment on the possible functional contribution $
S^{(1)}_{\GR\, \,\text{function}}$ in gravity. Assuming locality and
dimensional arguments again rules out any contribution here as well.

\section{Sub-subleading soft graviton operator}

The discussion for the sub-subleading soft operator for graviton amplitudes is
analogous to the subleading case. The starting point is an ansatz
for $\ssld_\GR$ of the form
\begin{equation}
  \label{eq:s2ansatz}
  \ssld_\GR = \sum_{a=1}^n E_{\mu\nu} \,\Omega_a^{\mu\nu\rho\sigma\gamma\lambda} \,
  \Lambda_{a,\rho\sigma} \, \Lambda_{a,\gamma\lambda}
\end{equation}
where we used $\Lambda_{a,\rho\sigma} := p_{a,\rho}\, \frac{\partial}{\partial
  p_a^\sigma} + E_{a,\rho}\, \frac{\partial}{\partial E_a^\sigma}$ as in \eqn{Lambdadef}. Again,
$\Omega_a$ must obey some constraints; specifically, it must have mass dimension
zero, it must vanish linearly in the limit $q\to 0$, it must be symmetric in the exchange
$\mu\leftrightarrow\nu$ and it must be symmetric in the simultaneous exchange
$\rho\leftrightarrow\gamma,\,\sigma\leftrightarrow\lambda$. The most general
ansatz satisfying these constraints is
\begin{equation}
  \label{eq:omegaansatz}
  \begin{aligned}
    \Omega_a^{\mu\nu\rho\sigma\gamma\lambda} = &
    \frac{\cff{1}}{(q\cdot p_a)^3} p_a^\mu p_a^\nu q^\rho q^\sigma q^\gamma q^\lambda +
    \frac{\cff{2}}{(q\cdot p_a)}  \eta^{\sigma(\mu} \eta^{\nu)\lambda} q^\rho  q^\gamma  +
    \frac{\cff{3}}{(q\cdot p_a)}  \eta^{\rho(\mu} \eta^{\nu)\gamma} q^\sigma  q^\lambda   \\
    & + \frac{\cff{4}}{(q\cdot p_a)} \bigl[\eta^{\rho(\mu} \eta^{\nu)\sigma} q^\gamma  q^\lambda  + \eta^{\gamma(\mu} \eta^{\nu)\lambda} q^\rho  q^\sigma   \bigr] +
    \frac{\cff{5}}{(q\cdot p_a)} \bigl[\eta^{\rho(\mu} \eta^{\nu)\lambda} q^\gamma  q^\sigma  + \eta^{\gamma(\mu} \eta^{\nu)\sigma} q^\rho  q^\lambda   \bigr]  \\
    & + \frac{\cff{6}}{(q\cdot p_a)^2} p_a^{(\mu} \eta^{\nu)(\rho} q^{\gamma)} q^\lambda q^\sigma +
    \frac{\cff{7}}{(q\cdot p_a)^2} p_a^{(\mu} \eta^{\nu)(\lambda} q^{\sigma)} q^\rho q^\gamma \ .
  \end{aligned}
\end{equation}
Furthermore $\ssld_\GR$ must obey the distributional constraint eqn.~\eqref{eq:cconstr2} and
the gauge invariance constraint on the soft leg. We recall these constraints
here
\begin{align}
  &\bigl[ \ssld_\GR, \delta^{(D)}(P)\bigr] = \tfrac{1}{2} \,S^{(0)}_{\GR}\, \Bigl(
  \bigl( q\cdot\partial \bigr)^2 \delta^{(D)}(P)\biggr) +  \Bigl(
  q\cdot\partial
  \delta^{(D)}(P) \Bigr) \, S^{(1)}_{\GR} + \chi' \, \delta^{(D)}(P)\,, \\
  &\ssld_\GR [E\to \Lambda q] \cdot \mathcal{A}_n = 0 \ .
\end{align}
Imposing these constraints yields a total of 5 linear equations for the seven
unknowns $\cff{j}$ residing in the ansatz \eqn{eq:omegaansatz} at each
leg\footnote{There is actually one additional equation that identifies the
  coefficients $\cff{4}$ in the ansatz with the undetermined coefficients
  $\half\tilde{c}^{(a)}$ appearing in the final form of $\sld_\GR$,
  eqn.~\eqref{eq:s1final}}.  A tedious but straightforward computation shows that
the solution is $\chi'=0$ and
\begin{align}
  \label{eq:s2res}
  \ssld_\GR = \frac{1}{2} \sum_{a=1}^n & \frac{1}{q\cdot p_a}
  E^{\lambda\sigma} q^\rho q^\gamma \, J_{a,\rho\sigma} \, J_{a,\gamma\lambda} + \nn \\
  +\sum_{a=1}^n &\biggl\{
  \frac{\tilde c_1^{(a)}}{q\cdot p_a}
  \biggl(\frac{(p_a\cdot E)(q\cdot E_a)}{q\cdot p_a} - E\cdot E_{a} \biggr )^{2}\, \biggl(q\cdot\frac{\partial}{\partial  E_a}\biggr)^{2}\nn \\ &
  + \tilde c_2^{(a)}\, \biggl(\frac{(p_a\cdot E)(q\cdot E_a)}{q\cdot p_a} - E\cdot E_{a} \biggr )\, 
  \biggl( \frac{E\cdot p_{a}}{q\cdot p_{a}}\, q\cdot\frac{\partial}{\partial  E_a} 
  - E\cdot\frac{\partial}{\partial  E_a} \biggr)\times \nn\\ &\qquad\qquad 
 \times\biggl(\frac{E_{a}\cdot q}{q\cdot p_{a}}\, q\cdot\frac{\partial}{\partial  E_a}
 + q\cdot\frac{\partial}{\partial  p_a}\, \biggr)
    \biggr\}
  \end{align}
%
% (Eq.dEa p.q - Eq.p q.dEa) (Ea.q q.dEa + p.q q.dp)
%
%
where for convenience we expressed the soft graviton polarization as $E_{\mu\nu}
= E_\mu E_\nu$ and $J_{a,\rho\sigma} := p_{a,\rho}\derpa{\sigma} +
E_{a,\rho}\frac{\partial}{\partial E_a^\sigma} - (\rho\leftrightarrow\sigma)$ as
before.  In the notation of the ansatz eqn.~\eqn{eq:omegaansatz} we have $\tilde
c^{(a)}_{1}= - \cff{4}+ \cff{6}$ and $\tilde c^{(a)}_{2}=\tilde c^{(a)}$ the undetermined
parameter of $\sld_{\GR}$.  

Again we see that the orbital part of the sub-subleading soft graviton operator
$\ssld_\GR $ is entirely determined. The polarization-dependent parts on the
other hand are now determined up to two numerical factors for every hard leg; as
already stated, the coefficient $\tilde{c}_2^{(a)}$ equals $\tilde{c}^{(a)}$,
where the latter are the undetermined coefficients appearing in the final form
of $\sld_\GR$. We thus have one additional free coefficient for each hard
leg. It is also worth noticing that the additional, polarization-dependent terms
are manifestly gauge invariant.

%%%%%%%%%%%%%%%%%%%%%%%%%%%%%%%%%%%%%%%%%%%%%%%%%%%%%%%%%%%%%%%%%%%%%%%%%%%%%%%%
\section{Four dimensions and spinor helicity formalism}
\label{sec:4D}

Let us now consider the four-dimensional case, where we can use the
spinor-helicity formalism\footnote{See
  e.g.~\cite{Henn:2014yza,Elvang:2013cua} for a textbook treatment.} 
  and obtain additional constraints from little-group
scalings. Those constraints are particularly easy to access in four dimensions,
because the Lorentz group factorizes into two parts acting on holomorphic and
antiholomorphic spinors respectively. 

%------------------------------
\paragraph{Gauge theory.}
Taking a positive-helicity soft gluon for concreteness, its polarization vector
can be expressed in terms of a holomorphic reference spinor $\mu_\alpha$ as
\begin{equation}
  \label{eq:softpol}
  E^{(\scriptscriptstyle{+})}_{\alpha\dot{\alpha}} = \frac{\mu_\alpha \tilde{\lambda}_{q,\dot{\alpha}}}{\abr{\mu\,\lambda_q}} \ .
\end{equation}
The ansatz for $\sld_\YM$ in spinor-helicity variables reads
\begin{equation}
  \label{eq:s1ym4d}
  \sld_\YM = \sum_{a=1}^n E^{(\scriptscriptstyle{+})}_{\alpha\dot{\alpha}}
  \biggl[
  \Omega_a^{\alpha\dot{\alpha}\beta} \derl{a}{\beta}
  + \bar{\Omega}_a^{\alpha\dot{\alpha}\dot{\beta}} \derlt{a}{\beta}
  \biggr]\,.
\end{equation}
In order to yield the correct mass dimension for $\sld_\YM$, the coefficients
$\Omega_a,\,\bar{\Omega}_a$ must be of mass dimension $-\half$. In addition
$\Omega_a$ carries helicity $\half$  and $\bar{\Omega}_{a}$ 
helicity $-\half$ on leg $a$. Moreover, terms where
the open index $\dot{\alpha}$ comes from $\tilde{\lambda}_q$ do not contribute
to $\sld_\YM$, as those terms vanish after contracting with the polarization
tensor eqn.~\eqn{eq:softpol}. Combining these constraints, the most general
ansatz reads
\begin{align}
  \Omega_a^{\alpha\dot{\alpha}\beta} = & \frac{\cff{1}}{\abr{a\,q}\sbr{a\,q}} \,\lambda_a^\alpha \lambda_a^\beta \tilde{\lambda}_a^{\dot{\alpha}} \ , \\
  \bar{\Omega}_a^{\alpha\dot{\alpha}\dot{\beta}} = &
  \frac{\cffb{1}}{\abr{a\,q}\sbr{a\,q}} \lambda_a^\alpha \ltil{\alpha}_a \ltil{\beta}_a +
  \frac{\cffb{2}}{\abr{a\,q}\sbr{a\,q}} \lambda_q^\alpha \ltil{\alpha}_a \ltil{\beta}_q +
  \frac{\cffb{3}}{\abr{a\,q}} \lambda_q^\alpha \se^{\dot{\alpha}\dot{\beta}} \ .
\end{align}
Gauge invariance on the soft leg implies that the operator obtained by the
substitution $E_q \to q$ in $\sld_\YM$ annihilates the amplitude. The resulting
operator is
\begin{equation}
  \label{eq:YMgauge4d}
  \sld_\YM [E_q \to q] = - \sum_{a=1}^n
  \biggl[
  \cff{1}\, \lambda_a^\beta \derl{a}{\beta} \,+\,
  \cffb{1} \, \tilde{\lambda}_a^{\dot{\alpha}} \derlt{a}{\alpha} \biggr] \ .
\end{equation}
In principle we can allow the above operator to be any operator annihilating the
$n$-point tree-level gluon amplitude. Since tree-level gluon amplitudes in four
dimension are invariant under conformal transformations, we could in principle
allow $\cff{1}=\cffb{1}=c$  for some fixed constant $c$ for all hard legs $a$, so that the above
operator is the dilation operator $\mathfrak{D}$ \cite{Witten:2003nn}
\footnote{Strictly speaking, what appears here is $\mathfrak{D}-2$; however, the
  constant piece could be in principle restored by adding a constant term
  $\sum_a E_{\alpha\dot{\alpha}} \frac{\lambda_a^\alpha
    \ltil{\alpha}_a}{\abr{a\,q}\sbr{a\,q}} $ to $\sld_\YM$. We will however see
  that, although allowed by gauge invariance, these terms are ruled out by the
  distributional constraint.}. % -- up to a constant shift that
% can be recovered by adding a constant term to $\sld_\YM$.

The distributional constraint eqn.~\eqref{eq:cconstr1} then reads
\begin{align}
  \label{eq:distr4d}
  \begin{aligned}
  \sum_{a=1}^n
  \biggl[
  2 c \frac{\abr{\mu\,a}}{ \abr{a\,q} \abr{\mu\,q}} \lambda_a^\alpha \ltil{\alpha}_a
   + (\cffb{2}+\cffb{3})  & \frac{1}{\abr{a\,q}} \lambda_a^\alpha \ltil{\alpha}_q
  \biggr]
  \frac{\partial}{\partial P^{\alpha\dot{\alpha}}} \delta^4 (P) \nn \\ &
  \overset{!}{=} 
  \frac{\abr{n\, 1}}{\abr{n\, q}\, \abr{q\, 1}}
  \biggl(\lambda_q^\alpha \ltil{\alpha}_q \frac{\partial}{\partial P^{\alpha\dot{\alpha}}} \delta^4 (P)\biggr)  + \chi \, \delta^4(P) \, ,
  \end{aligned}
\end{align}
where we have inserted the spinor-helicity form of $S_\YM^{(0)}$ on the r.h.s.
Since the first term in the l.h.s. cannot conspire to build a Lorentz generator, we
see that upon using Schouten's identity the solution to this equation
is\footnote{Notice that in $\sld_\YM$ only the combination $\cffb{2}+\cffb{3}$
  appears once $\bar\Omega_{a}$ is contracted with the polarization $E^{(+)}$, therefore we can consider this sum as a single coefficient.}
\begin{equation}
  \label{eq:sol4dYM}
  k=0 \ ,\qquad \chi = 0\ , \qquad \cffb{2}+\cffb{3} =
  \begin{cases}
    1\;\text{for }a=1,n \ ,\\
    0 \;\text{otherwise} \ .
  \end{cases}
\end{equation}
This leads to the known form for the four-dimensional subleading soft factor for gluon amplitudes,
\begin{equation}
  \label{eq:s1YMsol4d}
  \sld_\YM = \frac{\tilde{\lambda}_q^{\dot{\alpha}}}{\abr{q\,1}} \derlt{1}{\alpha} -
  \frac{\tilde{\lambda}_q^{\dot{\alpha}}}{\abr{q\,n}} \derlt{n}{\alpha} \ .
\end{equation}
%

%------------------------------
\paragraph{Gravity.}
For what concerns graviton amplitudes, we consider a positive-helicity soft
graviton; the polarization vector is expressed in terms of two reference spinors
$\lambda_{x}$ and $\lambda_{y}$ as
\begin{equation}
  \label{eq:polG}
  E_{\alpha\dot{\alpha}\beta\dot{\beta}}^{(\scriptscriptstyle{+})} :=
  \frac{1}{\abr{x\,q}\abr{y\,q}} \bigl(
  \lambda_{x,\alpha} \lambda_{y,\beta} + \lambda_{y,\alpha} \lambda_{x,\beta}
  \bigr)
  \tilde{\lambda}_{s,\dot{\alpha}} \tilde{\lambda}_{s,\dot{\beta}} \ ,
\end{equation}
We now consider the usual local first order ansatz for $\sld_\GR$
\begin{equation}
  \label{eq:s1ans}
  \sld_\GR = \sum_{a=1}^n E_{\alpha\dot{\alpha}\beta\dot{\beta}}^{(\scriptscriptstyle{+})} \Bigl[
  \Omega_a^{\alpha\dot{\alpha}\beta\dot{\beta}\gamma} \derl{a}{\gamma} +
  \bar{\Omega}_a^{\alpha\dot{\alpha}\beta\dot{\beta}\dot{\gamma}} \derlt{a}{\gamma}
  \Bigr] \, .
\end{equation}
Again $\Omega_{a}$ and $\bar{\Omega}_{a}$ must obey some constraints. The mass dimensions
are  $[\Omega_{a}]=[\bar\Omega_{a}]=\tfrac{1}{2}$, the helicity of the soft leg should be
zero for both $\Omega_{a}$ and $\bar{\Omega}_{a}$  and the helicities for leg $a$ are
$\half$ ($-\half$) for $\Omega_{a}$ ($\bar\Omega_{a}$). Moreover, the open indices
$\dot{\alpha},\,\dot{\beta}$ cannot come from $\tilde{\lambda}_q$ and both
$\Omega_{a}$ and $\bar{\Omega}_{a}$ must be symmetric in the pairs
$(\alpha,\beta)$ and $(\dot{\alpha},\dot{\beta})$ 
These constraints imply that the possible forms of $\Omega_a$ and
$\bar{\Omega}_a$ are
\begin{align}
  \label{eq:Wans}
    \Omega_a^{\alpha\dot{\alpha}\beta\dot{\beta}\gamma} = &
    \frac{\cff{1}}{\abr{a\,q}\sbr{a\,q}} \, \ltil{\alpha}_a \ltil{\beta}_a \lambda_a^\alpha \lambda_a^\beta \lambda_a^\gamma \ , \\
    \label{eq:Wbans}
      \bar{\Omega}_a^{\alpha\dot{\alpha}\beta\dot{\beta}\dot{\gamma}} = &
      \frac{\cffb{1}}{\abr{a\,q}\sbr{a\,q}} \, \ltil{\alpha}_a \ltil{\beta}_a \lambda_a^\alpha \lambda_a^\beta \ltil{\gamma}_a
      + \frac{\cffb{2}}{\abr{a\,q}\sbr{a\,q}} \, \ltil{\alpha}_a \ltil{\beta}_a \lambda_a^{(\alpha} \lambda_q^{\beta)} \ltil{\gamma}_q
  + \frac{\cffb{3}}{\abr{a\,q}}\, \se^{\dot{\gamma}(\dot{\alpha}} \tilde{\lambda}_a^{\dot{\beta})} \lambda_a^{(\alpha} \lambda_q^{\beta)} \ .
\end{align}
An infinitesimal gauge transformation amounts to the shift
\begin{equation}
  \label{eq:Gshift}
  \lambda_x \,\to\, \lambda_x + \eta \lambda_q\ , \qquad
  \lambda_y \,\to\, \lambda_y + \eta^{'} \lambda_q\ , 
\end{equation}
for some (infinitesimal) $\eta,\,\eta^{'}$.  Gauge invariance then implies that
\begin{equation}
  \label{eq:s1gauge4dsol}
  \cff{1} = \cffb{1} = 0 \ , \qquad \cffb{3}-\cffb{2} = c\, , \; \forall\, a
\end{equation}
for some universal constant $c$. $\sld_\GR$ then reads\footnote{Again, only the
  combination $\cffb{3}-\cffb{2}$ appears in $\sld_\GR$ once the contraction with $E^{(+)}$
  is performed.}
\begin{equation}
  \label{eq:s1grinterm}
  \sld_\GR = c \sum_{a=1}^n \frac{\sbr{q\,a}}{\abr{a\,q}} \frac{1}{\abr{x\,q}\abr{y\,q}}
  \, \bigl(\abr{a\,x}\abr{q\,y} + \abr{a\,y}\abr{q\,x} \bigr)
  \, \ltil{\gamma}_q \derlt{a}{\gamma}
\end{equation}
We can now impose the distributional constraint again,
eqn.~\eqref{eq:cconstr1}. This reads (after using Schouten's identity)
\begin{equation}
  \label{eq:s1distr4d}
  \begin{aligned}
  &\biggl(\frac{\partial}{\partial P^{\gamma\dot{\gamma}} } \,\delta^4(P)\biggr) \biggl\{
  \biggl[ \sum_{a=1}^n 
  2 k \frac{\abr{x\,a}\abr{y\,a}}{\abr{x\,q}\abr{y\,q}}\frac{\sbr{q\,a}}{\abr{a\,q}} \,\lambda_q^\gamma \ltil{\gamma}_q
  - \frac{k}{\abr{x\,q}\abr{y\,q}} \ltil{\alpha}_q \ltil{\gamma}_q \bigl(\lambda_x^\alpha \lambda_y^\gamma + \lambda_y^\alpha \lambda_x^\gamma \bigr) \sum_{a=1}^n \lambda_{a,\alpha} \tilde{\lambda}_{a,\dot{\alpha}}
  \biggr] \\
  \overset{!}{=}& \biggl(\lambda_q^\gamma \ltil{\gamma}_q \frac{\partial}{\partial P^{\gamma\dot{\gamma}} } \,\delta^4(P)\biggr)
  \sum_{a=1}^n \frac{\abr{x\,a}\abr{y\,a}}{\abr{x\,q}\abr{y\,q}}\frac{\sbr{a\,q}}{\abr{a\,q}}
  + \,\chi\,\delta^4(P) \ ,
  \end{aligned}
\end{equation}
where we wrote explicitly the form of $S_\GR^{(0)}$ in spinor-helicity
variables. Notice that the second term in the first line is zero in the 
distributional sense. We may therefore conclude that the solution is
\begin{equation}
  \label{eq:s14dsol}
  \chi = 0 \ , \qquad c = -\frac{1}{2} \ .
\end{equation}
This fixes the form of $\sld_\GR$ in eqn.~\eqref{eq:s1grinterm} to be
\begin{equation}
  \label{eq:s1grfinal}
  \sld_\GR = \frac{1}{2} \sum_{a=1}^n \frac{\sbr{a\,q}}{\abr{a\,q}}
  \biggl(\frac{\abr{a\,x}}{\abr{q\,x}} + \frac{\abr{a\,y}}{\abr{q\,y}} \biggr)
  \ltil{\gamma}_q \derlt{a}{\gamma} \ .
\end{equation}
In fact, we have also checked that the sub-subleading soft factor $\ssld_\GR$ in four
dimensions is completely fixed by gauge invariance and the distributional
constraint.

%%%%%%%%%%%%%%%%%%%%%%%%%%%%%%%%%%%%%%%%%%%%%%%%%%%%%%%%%%%%%%%%%%%%%%%%%%%%%%%%
\section{Discussion}

In this note we analyzed constraints arising for the novel subleading soft
gluon and graviton theorems in general dimensions.  Next to the obvious demands
of Poincar\'e and gauge invariance we pointed out a slightly less obvious
distributional constraint arising from the unbalanced arguments of the total
momentum conserving Dirac delta functions on both sides of the soft theorems.
The distributional constraint \emph{requires} the subleading soft operators to
be differential operators of degree one (subleading) or two
(sub-subleading) in the hard momenta and relates them to the leading Weinberg
soft pole function. 

In the $D$-dimensional case we started from an ansatz compatible with dimensional
analysis and soft momentum scaling. We demonstrated that the entity of those
constraints determines the subleading soft gluon and graviton differential
operators as well as the sub-subleading soft graviton differential operator up
to a single numerical constant for every leg. The undetermined constant is
related to derivatives with respect to polarizations. 
Arbitrary functions commuting with the delta distributions could be added to
these operators and are generally unconstrained. However, taking scalings and
mass dimension constraints into account assuming locality, there is nothing
which can be written down at tree-level. 

Specializing to the four-dimensional case and employing the spinor-helicity
formalism, the same line of arguments was shown to \emph{entirely} fix the
subleading differential operators. This can be traced back to the factorization
of the Lorentz group in four dimensions. Upon fixing a unitary gauge, however,
there might be similar arguments from little-group scalings in other
dimensions. 

The operators so determined match the forms established in the literature at
tree level. Given that our arguments are very general the question arises
whether they apply to loop amplitudes as well: certainly, Poincar\'e
invariance, gauge invariance as well as the distributional constraint
eqn.~\eqn{eq:conscond} continue to hold.  

However, in the loop scenario we have to consider at least four novel
circumstances, which are not reflected in our ans\"atze for the subleading soft
operators in eqns.~\eqn{ansatzomega}, \eqn{ansatzomega2} and
\eqn{eq:omegaansatz}.  First, the loop corrections may contribute to the
unconstrained functional parts of $S^{(1)}$ and $S^{(2)}$, as is in fact the
case in the one-loop corrections reported in \cite{Bern:2014oka,He:2014bga}.
Interestingly, the operator $S^{(2)}_{\GR}$ may also receive first-order
differential corrections which should be related to the functional corrections
to $S^{(1)}_{\GR}$ by the distributional constraint eqn.~\eqn{eq:cconstr2}.
Second, we construct our ans\"atze employing dimensional analysis to constrain
the possible terms. The dimensionality of the couplings, however, allows for
dimensionless quantities such as $\log(\frac{-\mu^{2}}{q\cdot p_{a}})$ or
$\frac{q\cdot p_{a}}{-\mu^{2}}$, which
so far have not been accounted for in our ans\"atze. These terms arise in the
IR-divergent one-loop corrections to the soft operators reported in
\cite{Bern:2014oka}. In fact, this also introduces contributions of the form
$(\log \se)$ in the soft momentum expansion.  Third, the loop-corrections may
not respect our central assumption of locality.  Fourth, for gauge theory the
leading soft factor $S^{(0)}_{\YM}$ receives loop corrections, which feed into
the subleading constraint equations.

After incorporating the issues pointed out in the last paragraph, the
distributional constraint might be of use in the future in order to constrain
possible loop corrections to soft theorems. While our work constrains the possible
forms of the subleading soft gluon and graviton operators, it would be desirable
to have a deeper understanding towards the origin of the soft theorems.

\subsubsection*{Acknowledgments}

We thank G.~Bossard, T.~McLoughlin, B.~Schwab and especially
C.~Vergu for important discussions.  JP thanks the Pauli
Center for Theoretical Studies Z\"urich and the Institute for Theoretical
Physics at the ETH Z\"urich for hospitality and support in the framework of a
visiting professorship. The work of MdL and MR is partially supported by grant
no.\ 200021-137616 from the Swiss National Science Foundation.

%%%%%%%%%%%%%%%%%%%%%%%%%%%%%%%%%%%%%%%%%%%%%%%%%%%%%%%%%%%%%%%%%%%%%%%%%%%%%%%%
\bibliographystyle{nb}
% \bibliography{bibliothek}
\bibliography{soft}

\end{document}